\documentclass[aps,twocolumn,showpacs]{revtex4}
\usepackage{amssymb,amsmath}
\usepackage{graphicx}
\usepackage{graphics}
\usepackage{epsfig}

\begin{document}

\title{High-energy effective theory for a bulk brane}

\author{Claudia de Rham$^{(1)}$,  Shunsuke Fujii$^{(2)}$,
Tetsuya Shiromizu$^{(2,3,4)}$ and Hirotaka Yoshino$^{(2)}$}


\affiliation{$^{(1)}$DAMTP, University of Cambridge, Wilberforce Road, Cambridge CB3 0WA, UK}

\affiliation{$^{(2)}$Department of Physics, Tokyo Institute of
  Technology, Tokyo 152-8551, Japan}


\affiliation{$^{(3)}$Department of Physics, The University of Tokyo,  Tokyo 113-0033, Japan}

\affiliation{$^{(4)}$Advanced Research Institute for Science and Engineering,
Waseda University, Tokyo 169-8555, Japan}

\date{\today}\vspace{5pt}

\begin{abstract}
We derive an effective theory describing the physics of a bulk brane
in the context of the RS1 model. This theory goes beyond the usual low
energy effective theory in that it describes the regime where the bulk
brane has a large velocity and the radion can change rapidly. We
achieve this by concentrating on the region where the distance between
the orbifold planes is small in comparison to the AdS length
scale. Consequently our effective theory will describe the physics
shortly before a bulk/boundary or boundary/boundary brane
collision. We study the cosmological solutions and find that, at large
velocities, the bulk brane decouples from the matter on the boundary
branes, a result which remains true for cosmological perturbations.
\end{abstract}

DAMTP-2005-90

\maketitle

\section{Introduction}

The study of brane collisions has recently gained a special interest as
it may provide a new scenario for the creation of the hot big-bang
Universe \cite{CU, Bubble}. Motivated by heterotic M-theory and the
Randall Sundrum (RS) model \cite{RSI}, the collision between two
orbifold branes has been explored, leading to a five-dimensional
singularity \cite{Khoury:2001bz}. When the boundary branes are close, an effective theory
can be derived for this scenario
and is hence valid just before or just after the collision \cite{collision, CL, CL2, CL3, CU}.
In this paper, we extend this analysis to the case where a brane is
present in the bulk. This regime is of interest as it allows us to
study a bulk/orbifold brane collision
in a situation where the five-dimensional
geometry remains regular similarly as in the first Ekpyrotic scenario
\cite{CU}. What is particularly interesting about the effective theory
we will develop, is that it is capable of describing the regime where
the branes have large velocities, something which the usual low energy
effective theory cannot do.



Although similar work has been derived for close boundary branes, \cite{CL, CL2,
CL3} it relied strongly on the presence of a
$\mathbb{Z}_2$ orbifold symmetry which is generically
broken for bulk branes.
This work will hence give us a general formalism for the derivation of an effective theory on a
generic non $ \mathbb{Z}_2 $-brane. Such branes are interesting to study as
they represent more realistic candidates for cosmology and at high-energies, their behaviour
is expected to be strongly modified \cite{Davis:2000jq}.

In order to get some insight on the brane geometry one should in
principal solve the full higher-dimensional theory exactly before
being able to infer the geometry on the brane. Unfortunately, this
is only possible in very limited cases, and for more general
situations, one should in practice either rely on numerical
simulation or work in some specific regime where effective theories
may be derived. This is the approach which is generally undertaken in
order to derive a {\it low-energy} effective theory. Assuming a
low-energy regime, it is possible to express the geometry on
the brane as the lowest order of a gradient expansion \cite{CU,GE,
Cotta, Moduli}. In this
paper, we use a similar method, but choose instead to work in a
 close-brane regime, where we only consider terms of leading order
 in the distance between the branes. This method allow us to
 highlight the presence of ``asymmetric" terms on the bulk brane
 (generic to the absence of $\mathbb{Z}_2 $-symmetry) which are
 negligible at low-energies and are usually discarded. As far as we
 are aware, 
this is the first effective-theory that models these
terms in a covariant way beyond the low-energy limit.\\
\indent The rest of this paper is organized as follows. In Sec. II, we
consider three branes and derive the effective theory on the
asymmetric bulk brane. In that theory, two scalar dynamical degrees
of freedom are present, namely the distance between the bulk brane
and each of the boundary branes. We point out the low-energy limit
of this theory, and check its consistency with previous results. In
particular, we show that the theory on the bulk brane is a standard
scalar-tensor theory of gravity coupled with two scalar fields.
In Sec. III, we apply our effective theory to cosmology and compare
our result with solutions from the five-dimensional theory. We show
that for large velocities, the matter on the orbifold branes do not
affect the bulk brane. As a specific example, we present the
derivation of tensor perturbations. As expected, at large velocities
the perturbations on the bulk brane decouple from the stress energy on
the boundary branes. Finally, we summarize our 
study and present some possible extensions as future works in Sec. IV.

\section{Effective theory for Three close branes}

Motivated by M-theory and the Randall-Sundrum model \cite{RSI}, we
consider spacetime to be effectively five-dimensional with 
the extra-dimension compactified on an $S^1/ \mathbb{Z}_2$-orbifold. Two
orbifold branes are located at the fixed point of the symmetry, and
we consider a third brane in the bulk.
In this paper, we shall be interested in the
limit where the three branes are close to each other, i.e.,
when they are either about to collide or have just emerged
from such a collision.
In this paper we use the index conventions that Greek
indices $\mu,\nu=0,\cdots,3$ are four dimensional, labeling the transverse $x^\mu$
directions, while Roman capital indices $M,N=0,\cdots,4$ are fully five dimensional
and lower cap Roman indices $i,j=1,2,3$ designate the spatial transverse
directions.
Without loss of generality, we use the
following metric ansatz
%
\begin{eqnarray}
ds^2=g_{MN}dx^M dx^N=e^{2\varphi_\pm(y,x)}dy^2+g_{\mu\nu}(y,x) dx^\mu dx^\nu
\end{eqnarray}
%
and we suppose that the branes are located at $y=y_+,y_0,y_-$.
The branes located at $y=y_\pm$ are the orbifold branes and they are
subject to a $ \mathbb{Z}_2 $-reflection symmetry.
The brane at $y=y_+$ is a positive tension brane, whereas the one at
$y=y_-$ has a negative tension. The brane located at $y=y_0$ is a
bulk brane and no symmetry is imposed.

In what follows, we denote by $\mathcal{R}_+$ (resp. $\mathcal{R}_-$)
the region between the bulk brane and the positive (resp. negative)
boundary brane. All through this paper we use the notation that an
index $'+'$ (resp. $'-'$) represents a quantity evaluated in the
region $\mathcal{R}_+$ (resp. $\mathcal{R}_-$), as shown in Fig. 1.
%
\begin{figure}
 \centering
 \includegraphics[width=.9\columnwidth]{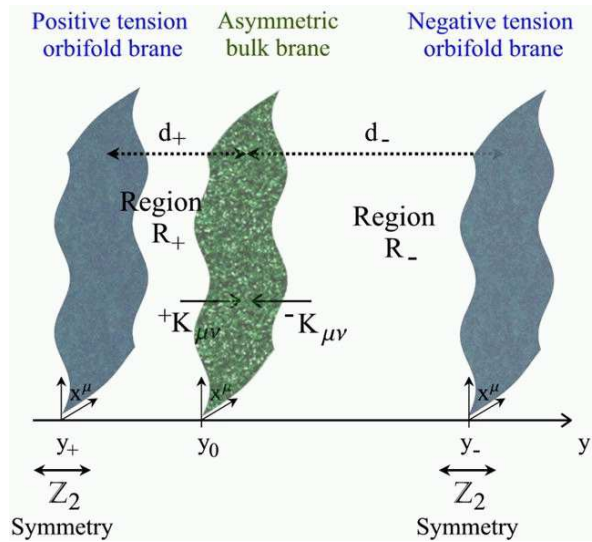}
 \caption{Two-brane Randall-Sundrum model with an asymmetric brane present in the bulk.}
\end{figure}
%
 In particular, the Anti-de Sitter(AdS) length scale on each region will be denoted
as $\ell_\pm$ and for any quantity $Q$, $^\pm Q(y_0)=\lim _{\epsilon
\rightarrow 0}Q(y_0\mp \epsilon)$. We denote by $^{(i)}T^\mu_\nu$
the stress-energy tensor for matter fields confined on the brane at
$y=y_i$, for $i=+,-,0$. We assume all branes to have a tension
$\sigma_i$ fine-tuned to their canonical value and absorb any
departure in their stress-energy: $\sigma_\pm=\pm 6/\kappa^2
\ell_\pm$ and $\sigma_0=3\left(\ell_-^{-1}-\ell_+^{-1}\right)/\kappa^2$
where $\kappa^2/8\pi$ is the five-dimensional gravitational constant.

The aim of this work is to derive an effective theory for the asymmetric
bulk brane. We will hence work on the bulk brane frame throughout
this paper unless otherwise specified. We will first decompose the
extrinsic curvature on the bulk brane, in terms of a quantity that may
be determined from the Isra\"el matching condition and another
``asymmetric'' quantity which needs to be determined by other
means. Working in the close-brane limit, we may express this
``asymmetric'' term as an expansion in terms of the extrinsic
curvature on the orbifold branes which are uniquely determined by the
junction conditions. The rest follows as in \cite{CL2, CL3}. In
particular, we
use the close-brane approximation to express the derivative of the extrinsic
curvature in terms of the extrinsic curvature on the bulk brane as
well as the one on the orbifold brane. This allows us to specify all unknown
quantities in the modified Einstein equation of the bulk brane and
hence obtain an effective theory for this brane.
%
%
\subsection{Expression of the ``asymmetric'' term}
For the boundary branes, the junction conditions are simply
%
\begin{eqnarray}
K^\mu_\nu(y=y_\pm) = - \frac{1}{\ell_\pm} \delta^\mu_\nu
\mp \frac{\kappa^2}{2} \Bigl({}^{(\pm)}T^\mu_\nu -\frac{1}{3}\delta^\mu_\nu
{}^{(\pm)}T \Bigr),
\end{eqnarray}
%
%
where $K^\mu_\nu$ is the extrinsic curvature.
Whereas for the bulk brane, due to the absence of any
$\mathbb{Z}_2$-reflection symmetry, the extrinsic curvature cannot
be uniquely determined by the junction conditions
%
\begin{eqnarray}
\Delta K^\mu_\nu(y_0) & := & \hspace{-5pt} {}^-K^\mu_\nu(y_0)-{}^+K^\mu_\nu (y_0)
\nonumber \\
\hspace{-5pt} & = & \hspace{-5pt}
\left(\frac{1}{\ell_+}-\frac{1}{\ell_-}\right)\delta^\mu_\nu
- \kappa^2 \Bigl({}^{(0)}T^\mu_\nu -\frac{1}{3}\delta^\mu_\nu {}^{(0)}T
\Bigr).\hspace{10pt} \label{junction bulk brane}
\end{eqnarray}
%

Since we are interested in the close-brane limit, we assume in
what follows, that the proper distance between the branes is much
shorter than the AdS curvature scale. In this case, the following
recursive relation is valid \cite{CL2,CL3}:
%
\begin{eqnarray}
{}^{\pm}K^{\mu (n)}_\nu (y_0)= \hat O_\pm K^{\mu (n-2)}_\nu
(y_0)(y_\pm-y_0)^{-2}, \label{rec}
\end{eqnarray}
%
where ${}^{\pm}K^{\mu (n)}_\nu (y_0)\equiv \partial_y^{\, n}\,{}^{\pm}\!K^{\mu}_\nu (y_0)$
and the
action of the operator $\hat O_\pm$ is defined by
%
\begin{eqnarray}
\hat O_\pm S^\mu_\nu =  D^\mu d_\pm D_\alpha d_\pm S^\alpha_\nu +
                          D_\nu d_\pm  D^\alpha d_\pm S^\mu_\alpha -(D d_\pm)^2
                          S^\mu_\nu \hspace{5pt}
\end{eqnarray}
%
for any symmetric tensor $S_{\mu\nu}=S_{(\mu\nu)}$.
The proper distance between the branes is
%
\begin{eqnarray}
d_\pm :=\pm e^{\varphi_\pm (y_0,x)} (y_0-y_\pm),
\end{eqnarray}
%
in the gauge where $\varphi$ is independent of $y$. In Ref. \cite{CL2},
it is shown that the result is independent of this gauge choice.

Working in such a gauge, and using the five-dimensional Einstein
equations, the Gauss equation on 
the brane is (Cf. Ref. \cite{SMS}),
%
\begin{multline}
{}^{\pm}R^\mu_\nu (y_0)  =   -\frac{4}{\ell_\pm^2}\delta^\mu_\nu
+{}^{\pm}K(y_0) {}^{\pm}K^\mu_\nu (y_0) \\
+ e^{-\varphi_\pm}{}^{\pm}K'^\mu_{\ \nu} (y_0)  +d_\pm^{-1}D^\mu D_\nu d_\pm. \label{ricci}
\end{multline}
%
%
The remaining task for the derivation of the effective theory is the
evaluation of both ${}^{\pm}K^\mu_\nu(y_0)$ and ${}^{\pm}K'^\mu_{\
\nu} (y_0)$. We hence decompose ${}^{\pm}K^\mu_\nu(y_0)$ into a
``known" contribution $\Delta K^\mu_\nu$, and an undetermined part
which represents the asymmetry across the brane:
%
\begin{eqnarray}
{}^{\pm}K^\mu_\nu (y_0) = \mp \frac{1}{2}\Delta K^\mu_\nu (y_0)
+\bar
K^\mu_\nu (y_0), \label{Kbulk}
\end{eqnarray}
%
where $\bar K^\mu_\nu (y_0) :=\frac 1 2 \left({}^{+}K^\mu_\nu (y_0)+
{}^{-}K^\mu_\nu (y_0) \right) $ and where the Codacci equation
holds for both quantities independently: $D_\mu \left(\Delta
K^\mu_\nu-\Delta K \delta^\mu_\nu\right) =D_\mu \left( \bar
K^\mu_\nu-\bar K \delta^\mu_\nu\right)=0 $. For a $ \mathbb{Z}_2$-symmetric
brane, $\bar{K}^\mu_\nu=0$.

Writing the extrinsic curvature on the orbifold branes as a Taylor
expansion in terms of the one on the bulk brane, we have:
%
\begin{eqnarray}
K^\mu_\nu (y_\pm) & = & \sum_{n \geq 0}^\infty \frac{1}{n!}{}^{\pm} K^{\mu (n)}_\nu (y_0)
(y_\pm-y_0)^n
\nonumber \\
& = & \frac{{\rm sinh}{\sqrt {\hat O}_\pm}}{{\sqrt {\hat
      O}_\pm}}{}^{\pm}K'^\mu_{\ \nu}(y_0) (y_\pm-y_0) \label{taylor}
\\ 
& & +{\rm cosh}{\sqrt {\hat O}_\pm}{}^{\pm} K^\mu_\nu(y_0).\nonumber
\end{eqnarray}
%
This provides an expression for the derivative of the extrinsic
curvature
%
\begin{multline}
{}^{\pm}K'^\mu_{\ \nu} (y_0)  =   \frac{1}{y_\pm -y_0}
\frac{{\sqrt {\hat O_\pm}}}{{\rm sinh}{\sqrt {\hat O_\pm}}}
\Biggl(K^\mu_\nu (y_\pm)  \label{derivative} \\
 -{\rm cosh} {\sqrt {\hat O_\pm}} {}^{\pm}K^\mu_\nu (y_0)
\Biggr).
\end{multline}
%
Substituting this expression into Eq. \eqref{ricci} and recalling that the
metric should be continuous across the bulk brane, ie.
${}^{+}R^\mu_\nu (y_0)-{}^{-}R^\mu_\nu(y_0)=0$, we obtain the
constraint
%
\begin{eqnarray}
&&\sum_{i=\pm}-\frac{i}{d_i}D^\mu D_\nu d_i+
\frac{4i}{\ell_i^2}\delta^\mu_\nu
-i\ {}^{i}\!K(y_0)\, {}^{i}\!K^\mu_\nu(y_0) \\
&&+
\frac{1}{d_i} \frac{{\sqrt {\hat O_i}}}{{\rm sinh}{\sqrt
{\hat O_i}}}
\Biggl(K^\mu_\nu (y_i) -{\rm cosh} {\sqrt {\hat O_i}} \
      {}^{i}K^\mu_\nu (y_0) \Biggr)=0. \nonumber 
\end{eqnarray}
%
Since the last two terms of the first line are of order $d^0$ and are hence
negligible compared to the other ones which are of order $d^{-1}$, we can solve the above
equation for the unknown part of ${}^{\pm}K^\mu_\nu (y_0)$:
%
\begin{eqnarray}
\bar K^\mu_\nu (y_0)= \hat L \ Z^\mu_\nu, \label{K bar}
\end{eqnarray}
%
where the operator $\hat L$ is
%
\begin{eqnarray}
\hat L = -\left[\frac{1}{d_-}\frac{{\sqrt {\hat O_-}}}{{\rm tanh}{\sqrt {\hat O_-}}}
+\frac{1}{d_+}\frac{{\sqrt {\hat O_+}}}{{\rm tanh}{\sqrt {\hat O_+}}} \right]^{-1}
\end{eqnarray}
%
and
%
\begin{multline}
Z^\mu_\nu  :=\sum_{i=\pm}\frac{i}{d_i}D^\mu D_\nu
d_i-\frac{1}{d_i}\frac{\sqrt{\hat{O}_i}}{\sinh \sqrt{\hat{O}_i}}\
K^\mu_\nu(y_i) \label{average}\\
-\frac{i}{2 d_i}
\frac{\sqrt{\hat{O}_i}}{\tanh \sqrt{\hat{O}_i}}\
\Delta K^\mu_\nu(y_0).
\end{multline}
%
It is worth pointing out that if a reflection symmetry was imposed
across the bulk brane, we would have $d_+=d_-$, $\hat O_+=\hat O_-$
and $K^\mu_\nu(y_+)=-K^\mu_\nu(y_-)$. The tensor $Z^\mu_\nu$
would hence vanish, and so would $\bar{K}^\mu_\nu$.\vspace{10pt}

%
\subsection{Effective-theory}
In the previous subsection, we have derived an expression for the
``asymmetric'' part of the extrinsic curvature in terms of quantities
that can be determined from the Isra\"el matching conditions and in
terms of the first derivative of the extrinsic curvature. 
As far as we are aware this is the first derivation of the ``asymmetric''
term beyond the low-energy regime. Knowing the extrinsic curvature on
the brane, we may use its expression in the Taylor expansion
\eqref{taylor}, to get an expression for the derivative
\eqref{derivative} which can finally be substituted into the modified
Einstein equation  \eqref{ricci}.

But first, we may express the equation of
motion for the radions, which can be derived from the traceless
property of the Weyl tensor ${}^{\pm}E^\mu_\nu (y_0)$. Using the
result of Ref.
\cite{CL,SMS}, the Weyl tensor 
can be formally expressed as
%
\begin{eqnarray}
{}^{\pm}E^\mu_\nu (y_0) = -e^{-\varphi_\pm}  {}^{\pm}K'^\mu_{\ \nu}
(y_0)
-\frac{1}{d_\pm}D^\mu
D_\nu d_\pm,
\end{eqnarray}
%
to leading order in $d_\pm$.
Since $E^\mu_\mu (y_0)=0$, this leads to the following Klein Gordon
equation for the distance between the branes:
%
\begin{eqnarray}
\hspace{-10pt}D^2 d_\pm =\hspace{-3pt}\pm\, \delta^\nu_\mu \frac{\sqrt{\hat{O}_\pm}}{\sinh
\sqrt{\hat{O}_\pm}}\Biggl[
K^\mu_\nu(y_\pm)\label{Box d}-\cosh \sqrt{\hat{O}_\pm}\, {}^{\pm}\!K^\mu_\nu(y_0)
\Biggr],
\end{eqnarray}
%
where the right-hand side can be computed from Eqs.
\eqref{Kbulk} and \eqref{K bar}.

The tracelessness of the Weyl
tensor together with the continuity constraint of the Ricci scalar
across the brane, also implies the supplementary constraint on the
asymmetric term $\bar K^\mu_\nu$:
%
\begin{eqnarray}
\Delta K \bar{K}-\Delta K^\alpha_\beta \bar{K}^\beta_\alpha
=6\left(\frac{1}{\ell_-^2}-\frac{1}{\ell_+^2}\right).
\label{HamConstraint}
\end{eqnarray}
%
The formal expression for the Ricci scalar on the bulk brane is therefore:
%
\begin{eqnarray}
{}^{(4)}R&=&-6\left(\frac{1}{\ell_+^2}+\frac{1}{\ell_-^2}\right)
+\bar{K}^2-\bar K^\alpha_\beta \bar{K}^\beta_\alpha \label{R}\\
&&+\frac 1 4 \Delta K^2-\frac 1 4\Delta K^\alpha_\beta \Delta
K^\beta_\alpha. \nonumber
\end{eqnarray}
%

We may now express the effective gravitational equation on the bulk brane. As
expected, it can be described by two scalar fields non-trivially
coupled to gravity:
%
\begin{multline}
{}^{(4)}G^\mu_\nu (y_0) = \frac{1}{d_\pm} (D^\mu D_\nu d_\pm -
\delta^\mu_\nu D^2 d_\pm)
\\ \mp
\frac{1}{d_\pm}
\left(\delta^\mu_\beta \delta^\alpha_\nu-\delta^\mu_\nu \delta^\alpha_\beta\right)
\frac{{\sqrt {\hat O_\pm}}}{{\rm
sinh}{\sqrt {\hat O_\pm}}}\\
\times
\Biggl[ K^\beta_\alpha (y_\pm)
-{\rm cosh}{\sqrt {\hat O_\pm}}{}^{\pm} K^\beta_\alpha (y_0)
\Biggr]. \label{G eff}
\end{multline}
The following formulae will be useful to rewrite the above equation in a more convenient way
%
\begin{eqnarray}
&\hat O \delta^\mu_\nu = 2D ^\mu d D _\nu d -\delta^\mu_\nu (D d)^2
\label{form1}\\
&\hat O^2 \delta^\mu_\nu =(D d)^4 \delta^\mu_\nu \label{form2}\\
&{\hat O}^2 S^\mu_\nu = 2D ^\mu d D _\nu d D ^\alpha d
D _\beta d
S^\beta_\alpha
-(D d)^2\hat O S^\mu_\nu
\label{form4}\\
&{\hat O}^{2n}S^\mu_\nu = (D d)^{4(n-1)} \hat O^2
S^\mu_\nu~~(n\geq 1)
\label{form5}\\
&{\hat O}^{2n+1}S^\mu_\nu = (D d)^{4n} \hat O S^\mu_\nu
\label{form6}
\end{eqnarray}
%
\begin{widetext}
%
\begin{eqnarray}
{\rm cosh}{\sqrt {\hat O}}S^\mu_\nu = S^\mu_\nu + \frac{{\rm
cos}|D d|-1 }{|D d|^2} \hat O S^\mu_\nu + \frac{{\rm
cosh}|D d|+ {\rm cos}|D d|-2}{|D d|^4}
 D ^\mu d D _\nu d D _\alpha d D ^\beta d S^\alpha_\beta 
\end{eqnarray}
%
%
\begin{eqnarray}
 \frac{{\sqrt {\hat O}}}{{\rm sinh}{\sqrt {\hat O}}} S^\mu_\nu
 =  S^\mu_\nu +\Biggl(\frac{|D d|}{{\rm sin}|D d|} -1 \Biggr)|D d|^{-2} \hat O S^\mu_\nu
+\Biggl(\frac{|D d|}{{\rm sin}|D d|} +\frac{|D d|}{{\rm sinh}|D d|}-2 \Biggr)
|D d|^{-4} D ^\mu d D _\nu d D _\alpha d
D ^\beta d S^\alpha_\beta 
\end{eqnarray}
%
%
\begin{eqnarray}
\frac{{\sqrt {\hat O}}}{{\rm tanh}{\sqrt {\hat O}}} S^\mu_\nu =
S^\mu_\nu +\Biggl(\frac{|D d|}{{\rm tan}|D d|} -1
\Biggr)|D d|^{-2} \hat O S^\mu_\nu
+\Biggl(\frac{|D d|}{{\rm tan}|D d|} +\frac{|D d|}{{\rm tanh}|D d|}-2 \Biggr)
|D d|^{-4} D ^\mu d D _\nu d D _\alpha d
D ^\beta d S^\alpha_\beta.
\end{eqnarray}
\end{widetext}
%
%
\subsection{Low-energy limit}
In the slow-velocity limit, the effective theory simplifies greatly. We
neglect the coupling of the radions to matter and neglect any terms
beyond second order in derivatives. In that case, the expression for
the asymmetric tensor $\bar{K}^\mu_\nu$ takes the form
%
\begin{eqnarray}
&&\hspace{-5pt}\bar K^\mu_\nu (y_0) =-\frac{1}{d}\left[d_-D^\mu
D_\nu d_+-d_+D^\mu D_\nu d_-\right]\\
&&\hspace{-5pt}-\frac{\kappa^2}{2d}\left[
d_- {}^{(+)}\tilde{T}^\mu_\nu-d_+ {}^{(-)}\tilde{T}^\mu_\nu
+\left(d_--d_+\right) {}^{(0)}\tilde{T}^\mu_\nu
\right]\nonumber\\
&&\hspace{-5pt}+\frac{1}{2\ell_+\ell_-}\left[
-\left(\ell_-+\ell_+\right)+\frac{1}{d}\left(\ell_+d_+\hat{O}_-+\ell_-d_-\hat{O}_+\right)
\right]\delta ^\mu_\nu , \nonumber
\end{eqnarray}
%
where $d:=d_++d_-$ and $\tilde{T}^\mu_\nu=T^\mu_\nu-\frac 1 3 T
\delta^\mu_\nu$. Using this expression in the modified Einstein
equation \eqref{G eff}, we obtain the induced Einstein tensor on the brane:
%
\begin{multline}
{}^{(4)}G^\mu_\nu (y_0)=\frac{1}{d}(D^\mu D_\nu d -\delta^\mu_\nu
D^2d )
+\frac{\kappa^2}{d}\ {}^{(\text{eff})}T^\mu_\nu\\
+\frac{1}{d}\Biggl[
\frac{1}{\ell_-} D ^\mu d_-  D _\nu d_-+\frac{1}{2\ell_-}\left( D
d_-\right)^2 \delta^\mu_\nu\\
-\frac{1}{\ell_+} D ^\mu d_+  D _\nu d_+-\frac{1}{2\ell_+}\left( D
d_+\right)^2 \delta^\mu_\nu
\Biggr],
\end{multline}
%
where
${}^{(\text{eff})}T^\mu_\nu=(1/2){}^{(+)}T^\mu_\nu+(1/2){}^{(-)}T^\mu_\nu+{}^{(0)}T^\mu_\nu$. 
This is precisely the close-brane limit of the low-energy theory
derived in \cite{Cotta}, and is hence a good consistency check.

The equations of motion for the two scalar fields can be derived
from Eqs. \eqref{Box d} and \eqref{HamConstraint}. Although
Eq. \eqref{Box d} appears as two different equations, they are not
independent and only give rise to the same following constraint for $ D^2 d$:
%
\begin{eqnarray}
D^2 d=\frac{\kappa^2}{3}\ {}^{(\text{eff})}T
+\frac{1}{\ell_-}\left(D d_-\right)^2
-\frac{1}{\ell_+}\left(D d_+\right)^2.
\end{eqnarray}
%
Using this result together with the continuity constraint
\eqref{HamConstraint}, we obtain the decoupled Klein-Gordon
equations for the two scalar fields at low-energy
%
\begin{eqnarray}
D^2 d_\pm=\frac{\kappa^2}{6}\left({}^{(\pm)}T\mp\frac{2\ell_\pm}{\ell_--\ell_+}{}^{(0)}T\right)
\mp\frac{1}{\ell_\pm}\left(D d_\pm\right)^2.
\end{eqnarray}
%
%
As another check, we can verify that this result is consistent with the
usual four-dimensional low-energy theory \cite{GE} if a reflection symmetry
was imposed across the brane. In that case $d_+=d_-$ and only one
scalar field is coupled to gravity.

%
\section{Applications}
In this section, we apply our effective theory
to cosmology and perturbations.
For the background solution, it is possible to solve the
Einstein equation exactly. We may therefore use this feature to
compare the exact five-dimensional result with our effective theory in
the close-brane limit. This provides us a useful check.
We will then use the effective theory in order to study cosmological
perturbations around this background.
%
\subsection{Cosmology}
%
%
\subsubsection{Five-dimensional solution}
In this subsection, we 
first solve the five-dimensional Einstein equation exactly assuming
cosmological symmetry (i.e. we assume the spacetime to be homogeneous
and isotropic along the three spatial directions tangent to the
branes). Working in the frame where the bulk is static, one may use
the Birkhoff's theorem to derive easily the exact form of the
solution. But in order to compare this solution with our effective
theory, it will be useful to work instead in the frame where the
branes are static. Such a change of frame is in general difficult to
perform, but working in the close-brane regime, and neglecting higher
order terms in the distance between the branes, the change of frame
may perform easily as it has been shown in Refs. \cite{CL2,CL3}. We
will hence use the result of these papers to infer the geometry on the brane.

In the frame where the bulk
is static, the geometry on both regions $\mathcal{R}_\pm$ is simply
Schwarzschild-Anti-de Sitter (SAdS) with black-hole mass parameter
$\mathcal{C}_\pm$:
%
\begin{eqnarray}
ds^2_{\mathcal{R}_\pm}&=&-n_\pm^2 dT_\pm^2+dY_\pm^2 +a_\pm^2 d\mathbf{x}^2\\
a_\pm^2&=&e^{-2 Y_\pm/\ell_\pm}+\frac{\mathcal{C}_\pm}{4}e^{2 Y_\pm/\ell_\pm}\nonumber \\
n_\pm^2&=&a^2_\pm-\frac{\mathcal{C}_\pm}{a_\pm^2}.\nonumber
\end{eqnarray}
%
It is important to notice that in this frame, the branes are not
static, as in the previous section, and we will assume the branes to
have loci $Y=Y_i(T)$. In particular, the bulk brane has loci
$Y_+=Y_0^{(+)}(T_+)$ with respect of the region $\mathcal{R}_+$, and loci
$Y_-=Y_0^{(-)}(T_-)$ as measured from an observed in the static bulk frame $\mathcal{R}_-$.
The induced line element on the bulk brane can be read off as:
%
\begin{eqnarray}
ds^2_{0}&=&-\left[n_0^2 - \left(\frac{d Y_0^{(\pm)}}{d
  T_\pm}\right)^2\right]dT_\pm^2+a_0^2 \, d\mathbf{x}^2\\
&=&-d t^2+a_0^2 \, d\mathbf{x}^2,
\end{eqnarray}
%
where $a_0$ is the induced scale factor on the bulk brane:
$a_0(t)=a_+\left(Y_+=Y_0^{(+)}(T_+(t))\right)=a_-\left(Y_-=Y_0^{(-)}(T_-(t))\right)$,
and similarly for $n_0$.
The physical time $t$ on the bulk brane, may be expressed in terms
of the five-dimensional time coordinate $T_\pm$: 
%
\begin{eqnarray}
d t^2 =\left[n_0^2 - \left(\frac{d Y_0^{(\pm)}}{d
  T_\pm}\right)^2\right]dT_\pm^2.
\end{eqnarray}
%
In order to derive the Friedmann equation on the branes, we may use the
Isra\"el junction conditions \eqref{junction bulk brane}
%
\begin{eqnarray}
\Delta K^i_j(y_0)=-\frac{\kappa^2}{3}\left(\sigma_0+\rho_0\right)
\delta^i_j,
\label{junction cosmology}
\end{eqnarray}
%
where $\rho_0$ is the energy density of matter fields located on the
brane and the extrinsic curvature is
%
\begin{eqnarray}
^{\pm}K^i_j(y_0)=\delta^i_j\
\left(1-\frac{\dot{Y}_0^{(\pm) \, 2}}{n_0^2}\right)^{-1/2}\left.\frac{d a_\pm(Y)}{a_\pm
d Y_\pm}\right|_{Y_0^{(\pm)}},
\end{eqnarray}
%
where $\dot{Y}_0^{(\pm)}=d Y_0^{(\pm)}(T_\pm)/d T_\pm$.  We may now re-express
the extrinsic curvature in terms of the Hubble parameter on the
brane. In particular we use the relation
%
\begin{eqnarray}
\dot{Y}_0^{(\pm)}=\frac{d a_0}{d t}\frac{d t}{d T_\pm}
\left(\left. \frac{d a_\pm(Y_\pm)}{dY_\pm}\right|_{Y_0^{(\pm)}}\right)^{-1}.
\end{eqnarray}
%
Using the fact that $d a_\pm/dY_\pm=-n_\pm/\ell_\pm$, we have:
%
\begin{eqnarray}
\dot{Y}_0^{(\pm)\, 2}=\frac{\ell_\pm^2 a^2_0 H^2}{1+ \ell^2_\pm \, a_0^2
  \, H^2/n_0^2}, \label{ydot}
\end{eqnarray}
%
where $H$ is the Hubble parameter on the brane $H =(da_0/dt)/a_0$. The
extrinsic curvature on each side of the bulk brane can therefore be
expressed in terms of the Hubble
parameter as:
%
\begin{eqnarray}
^{\pm}K^i_j(y_0)&=&-\delta^i_j\
\sqrt{\frac{n_0^2}{\ell_\pm^2 a_0^2}+H
^2},\notag \\
&=&-\delta^i_j\
\sqrt{\frac{1}{\ell_\pm^2}-\frac{\mathcal{C}_\pm}{\ell_\pm^2 a_0^4}+H
^2}.\label{K cosmology}
\end{eqnarray}
%
Having an expression for the extrinsic curvature in terms of the
Hubble parameter on each side of the bulk brane, we can therefore use the
Isra\"el matching condition Eq. \eqref{junction cosmology} to 
express the Hubble parameter on the brane in terms on the energy
density and the black-hole mass parameters. 
Substituting Eq. \eqref{K cosmology} into \eqref{junction cosmology}, we find the modified
Friedmann equation on the asymmetric bulk brane:
%
\begin{multline}
H ^2=\frac{T_b^2}{4}
+\frac{1}{4T_b^2}\left[\frac{1}{\ell_+^{2}}-\frac{1}{\ell_-^{2}}-\frac{1}{a_0^{4}}
\left(\frac{\mathcal{C}_+}{\ell_+^2}-\frac{\mathcal{C}_-}{\ell_-^2}\right)
\right]^2 \\
-\frac{1}{2}\left[\frac{1}{\ell_+^2}+\frac{1}{\ell_-^2}-\frac{1}{a_0^4}
\left(\frac{\mathcal{C}_+}{\ell_+^2}+\frac{\mathcal{C}_-}{\ell_-^2}\right)\right],
\label{Friedmann non Z2}
\end{multline}
%
with $T_b=\frac{\kappa^2}{3}\left(\sigma_0+\rho_0\right)$.
In this modified Friedmann equation, one might think that the
parameters $\mathcal{C}_\pm$ are arbitrary, but in 
what follows, we show that they depend strongly on the brane
velocities and find the precise relation between them. This is important
as it will allow us to compare this result with the one obtained from
the effective close-brane theory which gives a direct relation with
the brane velocities.

\subsubsection{Expression for the velocity of the branes.}

Without loss of generality, we work in the specific situation where
the bulk brane is about to collide with the positive boundary brane
($\dot d _+<0$),
and moves away from the negative boundary brane ($\dot d_->0$).
The velocity of both boundary branes should therefore be
positive while the velocity of the bulk brane should be negative.
In the case where the bulk brane has a positive canonical tension, $\ell_+>\ell_-$
the Hubble constant on the bulk brane will be positive.

We may now use the results of Refs. \cite{CL2,CL3} where 
the the following
relation for the radion velocity holds in the close-brane regime:
%
\begin{eqnarray}
\dot{d}_\pm= \mp
\left(\tanh^{-1}\left(\frac{v_\pm}{n_0}\right)+\tanh^{-1}\left(\frac{^{\pm}v_0}{n_0}\right)
\right),
\end{eqnarray}
%
where $v_\pm$ are the absolute value of velocities of the boundary branes with respect
to five-dimensional physical time at the collision and
$^{\pm}v_0=\left|d Y_0^{(\pm)}/d T_{\pm}\right|$
is the absolute value of the velocity of the bulk brane as measured by a static observer
in region $\mathcal{R}_\pm$. In what follows, a dot designates the
derivative with respect to the physical time $t$ on the bulk brane. 

Using the result of Ref. \cite{CL3}, one has:
%
\begin{eqnarray}
\left(\frac{v_\pm}{n_0}\right)=
\sqrt{1-\frac{1-\mathcal{C}_\pm/a_0^4}{\left(1\pm\frac16\kappa^2\ell_\pm\rho_\pm\right)^2}},
\end{eqnarray}
%
where $\rho_\pm$ is the energy density on the $\pm$-brane.
Using the expression \eqref{ydot}, one has the expression for the bulk
brane velocity
%
\begin{eqnarray}
\left(\frac{^{\pm}v_0}{n_0}\right)=\frac{\ell_\pm H }{\sqrt{\ell_\pm^2
H ^2+1-\mathcal{C}_\pm/a^4_0}}.
\end{eqnarray}
%
The radions' velocities
$\dot d_\pm$ can hence be expressed in terms of
the black-hole mass parameter $\mathcal{C}_\pm$
%
\begin{multline}
\dot d_\pm=\mp\tanh ^{-1}\sqrt{1-\frac{1-\mathcal{C}_\pm/a_0^4}{\left(1\pm\frac16
\kappa^2 \ell_\pm\rho_\pm\right)^2}}\\
\mp\tanh^{-1}\left(\frac{\ell_\pm
H }{\sqrt{\ell_\pm^2
H ^2+1-{\mathcal{C}_\pm}/{a_0^4}}}\right).
\end{multline}
%
We may use these equations to find an expression for the constants
$\mathcal{C}_\pm$ in terms of the brane velocities:
%
\begin{multline}
\frac{\mathcal{C}_\pm}{a_0^4}=\left[\frac{\pm\, \ell_\pm H +\sinh \dot d_\pm
\left(1\pm \frac16\kappa^2\ell_\pm\rho_\pm\right)}{\cosh \dot
d_\pm}\right]^2 \label{C and H} \\
 \mp\frac16{\kappa^2\ell_\pm}\rho_\pm\left(
2\pm\frac16{\kappa^2\ell_\pm}\rho_\pm
\right).
\end{multline}
%
Using this relation for $\mathcal{C}_-$, one has
%
\begin{eqnarray}
H^2+\frac{1-\mathcal{C}_-/a_0^4}{\ell_-^2}=
\left(\frac{1-\frac16{\kappa^2\ell_-}\rho_- +\ell_- H \sinh \dot
d_-}{\ell_-\cosh \dot
d_-}\right)^2. \label{H and C}
\end{eqnarray}
%
But from Eq. \eqref{Friedmann non Z2}, one has as well:
%
\begin{multline}
H^2+\frac{1-\mathcal{C}_-/a_0^4}{\ell_-^2}=\\
\left[
\frac{T_b}{2}-\frac{1}{2 T_b}
\left(\frac{1}{\ell_+^2}-\frac{1}{\ell_-^2}+
\frac{\mathcal{C}_-}{a_0^4\ell_-^2}-\frac{\mathcal{C}_+}{a_0^4\ell_+^2}\right)
\right]^2.
\end{multline}
%
Substituting the expression \eqref{C and H} for $\mathcal{C}_\pm$ into the last line, we
therefore have an equation for $H $ in terms of $\dot d_\pm$,
which has for solution:
%
\begin{eqnarray}
H =
\Lambda^{\text{eff}}+\frac16{\alpha \kappa^2}\rho_0^{\text{eff}},
\label{H and
dot d}
\end{eqnarray}
%
with the notation
%
\begin{align}
\rho_0^{\text{eff}}&=\frac{\rho_-}{2\cosh \dot
d_-}+\frac{\rho_+}{2\cosh \dot d_+}+\rho_0,\nonumber\\
\Lambda^{\text{eff}}&=\frac{\alpha}{2\ell_-}\left(1-\frac{1}{\cosh \dot
d_-}\right)
-\frac{\alpha}{2\ell_+}\left(1-\frac{1}{\cosh \dot d_+}\right), \nonumber\\
\alpha&=2 \left(\tanh \dot d_++\tanh \dot d_-\right)^{-1}. \nonumber
\end{align}
%
We may now compare this result with what is obtained from the
effective theory. But first, we might make some important remarks.
At low-energy, the contribution from the matter on each brane has an
equal weight: $\rho_0^{\text{eff}}\simeq
\left(\rho_-/2+\rho_+/2+\rho_0\right)$.
This is due to the fact, that
at low-energy, when the branes are close, the bulk geometry is
almost Minkowski and each brane has an equal contribution.
This result is however not obvious from the usual Friedmann equation (\ref{Friedmann non
Z2}), where only the matter on the bulk brane seems to contribute,
but one should take into account the expression of $\mathcal{C}_\pm$ in terms of $\rho_\pm$.
%
%
 However, at high-velocities, the situation is radically
different: The geometry on the bulk brane decouples entirely from
the  matter content on the orbifold branes (we may point out that
this result is valid for the two-brane case as well Cf. Ref. \cite{CL3}).
In that limit, we indeed have $\rho_0^{\text{eff}}\rightarrow \rho_0$, with an effective
cosmological constant $({1}/{2})\left|{1}/{\ell_+}-{1}/{\ell_-}\right|$.
Its contribution vanishes when the asymmetry
across the brane is maximal: $\ell_+=\ell_-$. In that case the bulk
geometry is almost unperturbed by the brane and the Friedmann
equation on the bulk brane couples quadratically to its own matter
 $H^2_0={\kappa^4}\rho_0^2/36$. This is an exact result
arising from the five-dimensional equations of motions in the close
brane and high-velocity limit. 
%
%
\subsubsection{Cosmology in the effective theory}
We now wish to compare this exact result with the predictions from
the close-brane effective theory.
The modified Einstein equation \eqref{G eff} on the
bulk brane reads:
%
\begin{eqnarray}
G^i_j(y_0)&=&\left(H ^2-\frac{1}{3}R^2\right)\delta^i_j \sim d^0 \nonumber\\
&=& \frac{1}{d_-}D^iD_jd_-+\frac{(y_--y_0)}{d_-}\, {}^-K^{\prime\,
i}_{\, \, j}\sim d^{-1}.\ \ 
\end{eqnarray}
%
Since the first line is of higher order in the distance between the
brane, the second line should vanish. The expression for the
derivative of the extrinsic curvature can be found in
\eqref{derivative}, and
using the relation $\hat{O}_\pm Z^i_j=\dot{d}_\pm^2 Z^i_j$ valid for
the background, we therefore have the equation: 
%
\begin{multline}
0=-H  \dot d_-\delta^i_j+\frac{\dot d_-}{\sinh \dot d_-}
K^i_j(y_-) \label{eq H}\\
-\frac{\dot d_-}{\tanh \dot d_-}\left(\frac{1}{2}\Delta
K^i_j(y_0)+\bar{K}^i_j(y_0)\right) ,
\end{multline}
%
where $\Delta K^i_j(y_0)$ is given in Eq. \eqref{junction cosmology},
%
\begin{eqnarray}
K^i_j(y_\pm)=\left(-\frac{1}{\ell_\pm}\mp
\frac{\kappa^2}{6}\rho_\pm\right)\delta^i_j, \label{Kij orbifold}
\end{eqnarray}
%
and $\bar{K}^i_j$ is given by:
%
\begin{eqnarray}
\hspace{-10pt}\bar{K}^i_j\hspace{-2pt}&=&\hspace{-2pt}-
\Biggl[
-\frac{\dot d_-}{d_-\sinh \dot d_-}K^i_j(y_-)
-\frac{\dot d_+}{d_+\sinh \dot d_+}K^i_j(y_+)\label{K bar
cosmology}\\
\hspace{-2pt}&+&\hspace{-2pt}\frac{1}{2}\left(\frac{\dot d_-}{d_-\tanh \dot d_-}-\frac{\dot
d_+}{d_+\tanh \dot d_+}\right)\Delta K^i_j(y_0)\notag\\
\hspace{-2pt}&+& \hspace{-2pt} H \left(\frac{\dot d_-}{d_-}-\frac{\dot
d_+}{d_+}\right)\delta^i_j\Biggr]
\left(\frac{\dot d_+}{d_+ \tanh \dot d_+}+\frac{\dot d_-}{d_- \tanh \dot
d_-}\right)^{-1}.\notag
\end{eqnarray}
%
Using these expressions \eqref{junction cosmology}, \eqref{Kij
orbifold} and \eqref{K bar cosmology} in the Eq. \eqref{eq H} for
$H $, we finally obtain the relation between the Hubble parameter on
the asymmetric
brane and the radions' velocities:
%
\begin{eqnarray}
H =
\Lambda^{\text{eff}}+\frac16{\alpha \kappa^2}\rho_0^{\text{eff}},
\end{eqnarray}
%
with the same notations as for Eq. \eqref{H and dot d}. This
corresponds precisely to what was obtained from the exact
five-dimensional theory and represents an important consistency
check.

We may also wonder whether this theory is capable of reproducing
the expression \eqref{Friedmann non Z2} for the Hubble parameter.
This Friedmann equation is a simple consequence of the tracelessness
of the Weyl tensor as we shall see. Using this property for the Weyl
tensor, we have indeed obtained in Eq. \eqref{R} an expression for
the Ricci scalar in terms of $\bar{K}^\mu_\nu$ and $\Delta
K^\mu_\nu$.  The expression of $\Delta K^\mu_\nu$ is in general
complicated, but for cosmological solutions the Eq. \eqref{HamConstraint} imposes
the constraint:
%
\begin{eqnarray}
\bar{K}^0_0=\left(-2-\frac{\Delta K^0_0}{\Delta K^1_1}\right)\bar{K}^1_1
+\frac{2}{\Delta K^1_1}\left(\frac{1}{\ell_+^2}-\frac{1}{\ell_-^2}\right),
\end{eqnarray}
%
which we may reexpress as
%
\begin{eqnarray}
\bar{K}^0_0=\frac{-\kappa^2p_0-3x}{x-\kappa^2\rho_0/3}\bar{K}^1_1
-\frac{2xy}{x-\kappa^2\rho_0/3},
\end{eqnarray}
%
where for simplicity we wrote $x=\ell_+^{-1}-\ell_-^{-1}$ and
$y=\ell_+^{-1}+\ell_-^{-1}$. Furthermore, from the Codacci equation,
we have the relation
%
\begin{eqnarray}
\partial_a \bar{K}^1_1(a)=a \left(\bar{K}^0_0-\bar{K}^1_1\right).
\end{eqnarray}
%
Using this result together with the conservation of energy condition
$p_0=-a\rho_0'(a)/3-\rho_0$, we may solve this
differential equation for $\bar{K}^1_1$ and obtain
%
\begin{eqnarray}
\bar{K}^1_1=-\frac{xy}{2\left(x-\kappa^2\rho_0/3\right)}
+\frac{C_{\text{A}}}{a_0^4\left(x-\kappa^2\rho_0/3\right)},
\end{eqnarray}
%
where the ``asymmetric" constant $C_{\text{A}}$ appears as an
integration constant. We may now use this expression in Eq. \eqref{R}:
%
\begin{eqnarray}
R&=&6aHH'(a)+12H^2 \nonumber\\
&=& \left[-\frac{\left({2C_{\text{A}}}/{a^4}-x y\right)^2}{2
T_b^3}+\frac 1 2 T_b^2\right]a\, \rho'_0(a)\\
&+&\frac{1}{T_b^2}\left[-12
\left(\frac{C_{\text{A}}}{a^4}\right)^2+\left(y^2-T_b^2\right)\left(x-T_b\right)\rho_0(a)
\right],\nonumber
\end{eqnarray}
%
where $T_b=T_b(a)=(1/3){\kappa^2}\left(\sigma_0+\rho_0(a)\right)$. This
is simply a first order differential equation for $H(a)$, of which
solution is
%
\begin{equation}
H^2=\frac{T_b^2}{4}+\frac{1}{4T_b^2}\left(
-xy+2\frac{C_{\text{A}}}{a_0^4}
\right)^2
-\frac 1 4 \left(x^2+y^2+\frac{C_S}{a_0^4}\right).
\end{equation}
%
The parameter $C_{\text{S}}$ appears as an integration
constant as well. This expression corresponds precisely to the Friedmann
equation \eqref{Friedmann non Z2} obtained by solving the
five-dimensional geometry, and we can now relate the black-hole mass
parameters to the integration constants $C_{\text{A}}$ and
$C_{\text{S}}$
%
\begin{eqnarray}
C_{\text{A}}&=&\frac{1}{2}\left(\frac{\mathcal{C}_+}{\ell_-^2}
-\frac{\mathcal{C}_-}{\ell_+^2}\right)\\
C_{S}&=&-2\left(\frac{\mathcal{C}_+}{\ell_-^2}+\frac{\mathcal{C}_-}{\ell_+^2}\right).
\end{eqnarray}
%
In particular, the contribution from $C_{\text{S}}$ should be
expected in a general case and is responsible for the dark energy
term. The contribution of $C_{\text{A}}$ is specific to the
asymmetric brane and should cancel when $\bar{K}^\mu_\nu=0$, this is
indeed the case when one has opposite AdS scales on each regions
$\mathcal{R}_\pm$: $\ell_-=-\ell_+$, and when the black-hole mass
parameters are the same: $\mathcal{C}_-=\mathcal{C}_+$. In a general
case it is not possible to deduce the expression of the asymmetric
term by solving the five-dimensional theory, and our theory provides
a useful alternative. This is for instance the case for the study of
perturbations.
%
%
\subsection{Cosmological perturbations}
The aim of this paper is to provide an effective theory capable of
describing a bulk brane geometry in a consistent way, beyond the
low-energy approximation.
In this paper we will hence not extend
this study to a large analysis of perturbations, which would be a
subject on its own, but we may point out some useful comments which
will be relevant for such a study. In particular, we have seen in the
previous section, that at high-velocities, the geometry on the bulk
brane decouples from the matter content of the orbifold branes. This
result was valid for cosmology and we may check the effect of
tensor perturbations.

First we may stress that the operators $\hat O_+$ and $\hat O_-$ do
not commute in general. For a symmetric tensor $S_{\mu \nu}=S_{\left(\mu
\nu\right)}$,
%
\begin{multline}
\left[\hat O_-,\hat O_+\right]S^\mu_\nu
= \Big[
\left(D^\mu d_-D_\beta d_+-D^\mu d_+D_\beta d_-\right)S^\beta_\nu  \\
+\left(D_\nu d_-D^\beta d_+-D_\nu d_+D^\beta d_-\right)S^\mu_\beta \Big]
\left(D_\alpha d_-\, D^\alpha d_+\right).
\end{multline}
%
So apart for cosmology and for tensor, vector perturbations or if we
work in a gauge where both $\delta d_-$ and $\delta d_+$ may be set
to zero, the
order of the action of the different operators in the effective
theory is important. In particular we expect this feature to have
consequences on the evolution of non-linear perturbations.

As a specific simple example, one might study here the evolution of
tensor perturbations:
%
\begin{eqnarray}
ds^2_0=a^2\left[-d\tau^2+\left(\delta_{ij}+h_{ij}\right)dx^i
dx^j\right],
\end{eqnarray}
%
where $\tau$ is the conformal time on the bulk brane and we write $h^i_j=\delta^{ik}h_{kj}$.
For tensor
perturbations, the right hand side of the modified Einstein equation
\eqref{G eff} is hence:
%
\begin{multline}
\delta G^i_j=\delta R^i_j=-\frac{({\dot d_+}/{2
d_+})X_-+({\dot d_-}/{2
d_-})X_+}{X_-+X_-}\ \dot{h}^i_j\\
+\kappa^2\, \frac{X_-X_+}{X_-+X_+}\ {}^{(\text{eff})}\!\delta
T^i_j ,
\end{multline}
%
with the effective matter contribution:
%
\begin{eqnarray}
\hspace{-15pt}{}^{(\text{eff})}\!\delta T^i_j=
\frac{1}{2 \cosh \dot d_-}\,{}^{(-)}\!\delta T^i_j
+\frac{1}{2 \cosh \dot d_+}\,{}^{(+)}\!\delta T^i_j
+\,{}^{(0)}\!\delta T^i_j,
\end{eqnarray}
%
where $^{(k)}\!\delta T^i_j$ is the tensor part of matter
perturbations on the brane at $y=y_k$.
For simplicity, we wrote $X_\pm=\dot d_\pm /\left(d_\pm \tanh \dot
d_\pm\right)$. The only non-negligible part in the perturbation of the
Ricci tensor is $\delta R ^i_j=-\frac{\bar{\partial^2}}{2 a_0^2}\, h^i_j$,
the evolution of the tensor perturbations, is hence controlled by
%
\begin{equation}
\hat{\boxdot} \, h^i_j=-a_0^2\,\Omega\kappa^2\ {}^{(\text{eff})}\!\delta T^i_j,
\end{equation}
with
\begin{align}
\hat{\boxdot}&=
\left[\bar{\partial^2}
-\frac{d'_+d'_-}{d_+d_-}\frac{\tanh \dot d_++\tanh \dot d_-}
{\frac{d'_+}{d_+}\tanh \dot d_++\frac{d'_-}{d_-}\tanh \dot
d_-}\partial_\tau
\right],\nonumber \\
\Omega&=2\left(\frac{d_+ \tanh \dot d_+}{\dot d_+}
+\frac{d_- \tanh \dot d_-}{\dot d_-}\right)^{-1},\nonumber
\end{align}
%
where a prime designates derivative with respect to
the conformal time $\tau$ and the operator $\bar{\partial^2}$ is the
Laplacian in Minkowski space. One may note that in the close brane
limit, if $d_\pm \sim \dot d_\pm t$, the damping term
$\frac{d'_+d'_-}{d_+d_-}\frac{\tanh \dot d_++\tanh \dot d_-}
{({d'_+}/{d_+})\tanh \dot d_++({d'_-}/{d_-})\tanh \dot d_-}$
simply goes as $\tau^{-1}$, which is what is expected from a
usual four-dimensional theory. The expression for the effective
four-dimensional Newtonian constant is on the other hand slightly
affected: $\kappa_{(4d)}^2=\kappa^2\Omega$, which is similar to the
result obtained in Ref. \cite{CL3}. For more sophisticated analysis,
we however expect the result to be more interesting, especially
when the operators $\hat O_\pm$ do not commute.

However, we may point out that the remarks formulated for the
background remain valid at the level of perturbations. Namely, for
large brane velocities, the effective matter contribution on the
bulk brane is
%
\begin{eqnarray}
{}^{(\text{eff})}\!\delta T^i_j\rightarrow\,{}^{(0)}\!\delta
T^i_j,
\end{eqnarray}
%
and perturbations are not sensitive to the matter content of
the orbifold branes.
This provides a braneworld scenario, where the branes
could be close (and hence the Kaluza Klein modes difficult to
excite and to affect the brane), and yet the geometry on the bulk
brane would decouple from the other ones. \vspace{5pt}

%
\section{Summary and Discussion}

In this paper, an effective theory describing the gravitational
behaviour of a bulk brane has been derived. The absence of any
reflection symmetry across a generic bulk brane makes its behaviour
especially interesting to study. In the ``light"-brane limit, ie.
when the five-dimensional geometry is almost unaffected by the
presence of the brane, the asymmetry on the brane itself is
important and affects its own behaviour. In this work, we have
developed a four-dimensional effective theory capable of describing
this ``asymmetry" in a covariant way. For that, we have
considered a close-brane approximation, where we assumed the bulk
brane to be close to both orbifold branes.

Using this approximation, we obtained a resulting theory of gravity coupled in
a non-trivial way with two scalar fields representing the distance
between the bulk brane and each of the orbifold branes. This
four-dimensional theory can be tested in several limits, such as at
low-energy, when a reflection symmetry is imposed by hand and for
cosmology. In all these regimes, predictions from the close-brane
theory agree perfectly with the expected results. The case of
cosmology is of special interest, at high-velocity the bulk geometry
is not sensitive to the matter present on the orbifold branes, and
this result remains valid for tensor perturbations. In the limit
where the AdS length scale is the same on both side of the brane,
ie. the bulk is not perturbed by the brane, we show that the Hubble
parameter couples linearly to the energy density on the bulk brane.
This is an interesting result, which might strongly affect the
gravitational behaviour on such a brane.

This effective theory could as well be derived on the orbifold
branes in the presence of such a brane in the bulk. We may point out
that the asymmetric tensor we derived on the bulk brane depends on the bulk
brane metric. It will hence be necessary to find its expression in
terms of the orbifold brane metric before being able to derive an
effective theory for these orbifold branes. This is left for a future study.

A straightforward extension of this model, would be the scenario where the bulk brane
is close to only one of the orbifold branes, and the radion
representing the distance with the other orbifold brane is moving
slowly. One side of the theory would hence be modeled by the
low-energy effective theory while the close-brane theory would be a
good description for the other side. Such a model would be of
interest if one considers the collision of the bulk brane with one
of the orbifold branes. Such a process might produce a phase
transition which could have some interesting consequences from a
cosmological point of view. This is also left for a future study.


\section*{Acknowledgements}

The work of TS was supported by Grant-in-Aid for Scientific
Research from Ministry of Education, Science, Sports and Culture of
Japan(No.13135208, No.14102004, No. 17740136 and No. 17340075).
CdR is supported by DAMTP and was invited by a FGIP visiting
program. CdR wishes to thank TITECH for its hospitality.

\end{document}